\documentclass[letterpaper]{article}
\usepackage{natbib,alifeconf}

\title{Emergence-focused design in complex system simulation}
\author{Chris Marriott$^{1}$ \and Jobran Chebib$^{2}$ \\
\mbox{}\\
$^1$University of Washington \\
$^2$University of Z\"urich}

\begin{document}
\maketitle

\begin{abstract}

Emergence is a phenomenon taken for granted in science but also still not well understood.  We have developed a model of artificial genetic evolution intended to allow for emergence on genetic, population and social levels.  We present the details of the current state of our environment, agent, and reproductive models.  In developing our models we have relied on a principle of using non-linear systems to model as many systems as possible including mutation and recombination, gene-environment interaction, agent metabolism, agent survival, resource gathering and sexual reproduction.  In this paper we review the genetic dynamics that have emerged in our system including genotype-phenotype divergence, genetic drift, pseudogenes, and gene duplication.  We conclude that emergence-focused design in complex system simulation is necessary to reproduce the multilevel emergence seen in the natural world.  

\end{abstract}

\section{Introduction}

Many of life's mysteries are systems resulting from the interactions between heterogeneous components in a lengthy flow of behavior, yet the results can be predicted due to linear interactions.  A mechanical clock is a great example and Rube Goldberg perfected this idea in art with his famous machines \citep{WG00}.  Such systems may be \emph{complicated} (i.e. have many different components) but they are not \emph{complex} (i.e. system behavior cannot be explained in terms of the behavior of the system's sub-systems).

Other mysteries of life are systems resulting from the interactions between homogeneous or heterogeneous components interacting such that the overall behavior is more than just the sum of its parts \citep{G87, J02}.  Such systems are called complex or non-linear.  Herding is an example of a complex population level behavior emerging from the simple behaviors of homogeneous agents \citep{R87, I01, OS06}.  From this example and many like it in other domains, complex systems science has itself emerged as we have begun to understand the nature of emergence.

Emergence, as it is currently understood, plays a privileged role in the current organization of science \citep{M02, E06}.  As \cite{F74} famously argued, while we commonly think of psychology, biology, chemistry and the like to be reducible to physics (technically the respective subjects of these studies), it turns out that no reduction exists, or even could exist.  From this perspective the objects of one science, like molecules in chemistry, are distinct kinds of entities from another, say atoms and subatomic particles in physics, and the behavior of one cannot be explained simply in terms of the behavior of the others. 

\cite{K92} helped bring some metaphsyical peace to reductionism by helping to explain how on the one hand everything \emph{is} physics, but on the other hand no clear reduction exists from one level of science to others.  The concept introduced was \emph{supervenience} the idea that the higher order objects and processes were the function of the organization of the underlying objects and processes.  Supervenience allowed for a continued commitment to a metaphysical physicalism about science.  However it raised an important question, namely, how did these higher-level objects come to be out of the lower-level ones?  Today this process is called \emph{emergence} (we refer to what is sometimes called \emph{weak emergence} \citep{C06}).   

If this story is accurate, then the world of chemistry emerged out of physics, and the world of biology emerged out of chemistry sometime later, and now the worlds of psychology, culture, economics, politics, art, and many others have emerged out of biology.  The objects of each level organize themselves out of the interacting elements at lower levels.  \cite{D95} calls this process of increasing complexity ratcheting (here we use the term scaffolding).  While the examples above have focused on broad levels of emergence, \cite{SM04} argue this type of emergence has occurred many times within biological evolution. 

Analytical science has made some progress attempting to handle emergence. This is especially true in the field of evolution.  Many analytical models have been developed to predict evolutionary trajectories and their accuracies have been successfully verified in simulation \citep{HW01, GO12, MM15}. Although analytic models built on understanding evolution in morphometric space may be useful in answering some specific questions, it is not clear that they are still predictive for multi-level complex systems, where fitness landscapes may be non-continuous \citep{P08}, dynamic, or dependent on the phenotypic distributions of the populations of interest \citep{NS04}.  Thus a primary tool for studying the dynamics of complex systems in evolution and elsewhere is computational simulation \citep{CR89, MP08, HHC09}.

Modern simulations of complex systems have led to many breakthroughs in understanding the mechanisms of emergence.  However in most of these simulations the same simple methodology is used.  In simulation the phenomena of study is treated as emergent, so the underlying phenomena is modeled as a system of interacting parts \citep{E99}.  The problem is that for simplicity the underlying system itself is commonly simulated as a linear system (simulations of non-linear dynamics all the way down to the physical level are not computationally feasible).  However, what is observed in nature is layers and layers of complex emergent systems.

It is our desire to study emergence of complex adaptive systems at multiple levels and in particular we are interested in the co-evolution of culture and biology in humans and other organisms.  We believe that one necessary next step in complex system simulation is to incorporate more complicated (heterogeneous, multi-level) component systems and have these component systems emerge from non-linear dynamics whenever possible.  We call this \emph{emergence-focused design}.

For our purposes we want to study the emergence of culture.  Commonly agent-based models choose to model agents in culture as simple linear systems (\cite{R87, B02, MW02}).  Our desire is to start at least one level lower so we begin with agents with behaviors determined by rich genetic-phenotypic interactions.  Our first attempt at multi-level emergence showed promise \citep{MC14} and so we have extended our model to include a structured environment.  This is our first step towards multi-level emergence (higher order emergence from a self-organized, emergent system) and we present the current state of our model.

\section{Model}

Our model has incorporated several properties that enrich the dynamics of evolutionary simulation.  First, our agents exist in a \emph{structured environment} following a principle that complexity in an agent must reflect complexity in its environmental interaction.  Our agents live in a random geometric network of foraging sites.  Sexual reproduction requires two agents to be in the same place at the same time which creates a complex social interaction problem.

Second, our agents have a \emph{structured genome} following a principle that genetic dynamics emerge from the physical interaction of an actual genetic sequence.  Each agent has a genome represented by a path of foraging attempts at sites in the network.  

Third, the genome must interact with the environment to express a phenotype following a principle of decoupling the phenotype from the genotype (a.k.a. ontogenetic development and learning).  The path of foraging sites in the agent's genome together with the current site of the agent determine the agent's daily behavior.

Fourth, the agent's survival and reproductive fecundity are functions of their interaction with their environment following a principle of \emph{natural selection} as opposed to artificial selection by a fitness function.  A day's foraging costs the agent energy and this energy is replenished by consuming the resources gathered.  This constitutes a simple \emph{metabolism} for each agent.  If an agent has a daily deficit of energy, it will eventually lose all of its energy and die.  If an agent has a daily surplus of energy, then it will store any extra energy over the daily maximum until it can reproduce.  

Each of these principles is intended to take a component of the evolutionary system that is often modeled as a linear system and allow it instead to have the potential for emergent dynamics.

\subsection{Environment}

\begin{figure}[t]
\begin{center}
\includegraphics{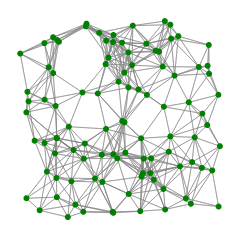}
\caption{A typical random geometric network with $n = 100$ sites and a connection reach of $0.2$.}
\label{fig:map}
\end{center}
\end{figure}

A random geometric network is created by randomly generating $n$ points on a plane and then connecting the close points \citep{P03, ABT14}.  For our network we generated points in the range $[0,1)\times[0,1)$ and connected points that had a Euclidean distance of at most $0.2$.  Figure~\ref{fig:map} shows a typical map generated in this method.  With these settings we usually get a connected network.  

Each site in the network has a foraging task for gathering resources.  We extend the model of \cite{MC14}.  Each site will have a fixed resource reward for any agent that forages at the site.  For typical settings our sites may have a reward of one, two or three resources.  Every agent receives the same reward at a site unless the site is depleted for the day in which case the foraging agent receives no reward.  Sites become depleted after a fixed number of foraging attempts at the site (we call this parameter the site's yield). 

Foraging at a site will cost an agent energy.  Energy in our model is temporal energy so corresponds to an amount of time spent on activities.  The energy cost is determined by the degree of success of the agent's foraging strategy at the foraging task.  The foraging task is to find the rewards that are hidden in five possible locations and the foraging strategy is the order in which hidden locations are inspected.  For example, a site with three rewards assumes these three rewards are hidden among five different locations (the number of hiding locations and the number of rewards are parameters that can be varied).  The agent must select an order to inspect the hidden locations and each check takes one unit of energy.  So the agent can find the three rewards after three tries, after four tries or after five tries.  The minimum energy cost at any site will be one unit of energy per resource gained.  The maximum will be the number of hiding locations.  The task at a particular site is currently held static throughout the simulation so as to be the target of optimization.  Mutations to strategies plus selection will allow for agents to adapt to the foraging tasks.

Moving in the environment also costs the agent energy.  The cost is proportional to the length of the edge they travel to get to the new site.  In our simulation agents are not allowed to forage at the same site twice in a row, though they may return if they forage at at least one other site first.  This is because their foraging behaviors are always represented as a \emph{path} in the network.

The total energy that an agent may spend in a day cannot exceed its current energy reserves or the maximum daily energy expenditure.  This means that agents with less than the maximum energy are free to use all of their energy to gather resources.  Agents with more than the maximum energy can only spend up to this maximum but the remaining energy can only be stored for reproductive purposes.  Energy is replenished at the end of the day by consuming the resources gathered.  Each resource consumed generates two energy for the agent (this rate can be varied).  In addition to foraging, breeding and travel energy costs, an agent is also penalized daily based on its age.  This penalty is a quadratic function of age growing faster as the agent gets older.  This ensures agents will eventually die even if they are well adapted to their environment.

At the moment we only test static environments so these features and parameters do not change over the course of the simulation.  However, because of the depleted resource sites the agents do experience some minor variation in their environment over the course of the day and between days.

\subsection{Agent}

The agents in our simulation have their behaviors determined by the interaction of their genomes with the environment.  The genome of an agent is represented by a path of foraging attempts at sites in the network.  This path in the network is a single unbroken path though it may and often does loop back on itself repeating sites.  We consider each site in this path a gene in the agent's genome. 

A seed agent for our simulation has a randomly generated genome.  We build the path by simulating a random walker on the network starting at a random site \citep{l93}.  At each site we must select a random foraging strategy for the agent to use.  This means selecting an order in which the agent will search the hidden locations at the foraging site.  We generate a random strategy for the site by selecting a permutation of the five hiding locations uniformly.  The gene for this site includes two other components in addition to the foraging strategy.  One component determines whether an agent will attempt to breed at the site, and if so, how much energy will be expended in this attempt.  This component is randomly generated for the seed agent.  Second, the agent must then travel to the next site, and the gene will have an energy cost associated with this travel cost.  This cost is fixed by the environment.  In this way we can see each gene, if acted upon, will have a total energy cost to the agent equal to the foraging cost plus the breeding cost plus the travel cost.  We can then use this metric to measure the energy cost of subpaths of the genome or of the genome as a whole.  When randomly generating a genome for a seed agent we generate one with total length about four times greater than the maximum daily energy expenditure.  With typical settings this is on average 54 genes.

Once a seed agent is generated we select a gene from its genome at random and start the agent in the site indicated by that gene.  Then the agent must select its actions for the day.  An agent selects its daily activities by searching its genome forwards and backwards for a subpath beginning at its current site and that maximizes its expected resource gain.  The length of the subpath must be less than or equal to the energy the agent has to spend on foraging this day.  The optimal path that is found then becomes the behavior of the agent for that day.  Since it searches both backwards and forwards (and allow subpaths to rebound off the ends of the genome) there is almost always more than one possible course of action for a day.  The agent will always select the optimal option.  Also since it can search backwards and forwards, the agent can always follow the optimal path it found yesterday again today, but in reverse order.  In this way the agents typically find locally optimal subpaths of their genome to repeat day after day.  This subpath is locally optimal, but may not represent the globally optimal subpath in the genome.

\subsection{Reproduction}

An agent can reproduce if it has stored enough energy.  How much energy is necessary is determined by what type of reproduction the agent engages in.  First, the total cost of reproduction is equal to the daily maximum energy expenditure.  This is because the parent agent(s) create a new agent with maximum energy and so observing a principle of conservation of energy the parent agent(s) must transfer that energy from their storage.  This means asexual reproduction is twice as expensive to the parent as sexual reproduction, though in both cases the offspring costs the same overall.  Agents will only engage in reproduction if their energy bank is enough to afford the cost of reproduction and leave them with at least the maximum daily energy expenditure.  With these energy dynamics it means that asexual reproduction is much rarer than sexual reproduction in cases when agents can find sexual partners.  In some simulations, asexual reproduction is disabled, though it is necessary when simulations begin with a single seed agent. 

Sexual reproduction is carried out between two agents.  For this to occur they must both be spending time seeking a mate at the same foraging site for overlapping time periods during the day.  This may seem a rare occurrence but due to forces of common descent and convergence the agents develop social organizations like herding, natal philopatry, and assortative mating, and so have no problem finding a mate in our simulations \citep{MC15}. 

When a new agent is created from reproduction, its new genome may mutate, recombine with itself, or recombine with another genome (in the case of sexual reproduction).  The mutation rate is set at five percent but can be altered.  In the simulation there are three kinds of random alterations of the genome.  

First, each gene has a mutation probability.  Specifically, the foraging strategy in the gene (the permutation) is allowed to mutate.  A permutation mutates by swapping two of its entries.  This may change the energy cost of that gene.  We consider each possible strategy at the same site a different allele of the same gene.  This type of mutation is the smallest type of change and it is not used in our current measurements of genetic difference.

Second, the path at either end of the genome may grow with some probability.  When this occurs, a random walker is simulated as described for the seed agent above.  The random walk begins at the site indicated by the last (or first) gene in the genome.  The length of growth is bound by a constant (this is a simulation parameter usually set to 20 genes).

Third, the path at either end of the genome may shrink with some probability.  This is done by selecting a length and an end at random and deleting that many genes from the path (the length of the deletion is bound by the same parameter as growth of a genome).
%
%

Recombination, another type of mutation that is less random, also has three forms: cycle copy, cycle deletion, and sexual recombination.  When an agent mutates it has equal chance of a random growth or deletion (as above) or a cycle copy or cycle deletion.  That is, when a mutation occurs, one of a random growth, a random deletion, a cycle copy or a cycle deletion occurs with uniform probability.  Thus, half the time the genome gets longer and half the time it gets shorter.  Half the time this is due to random mutation and half the time it is due to recombination.  

\begin{figure}[!t]
\begin{center}
\includegraphics[width = 0.5\textwidth]{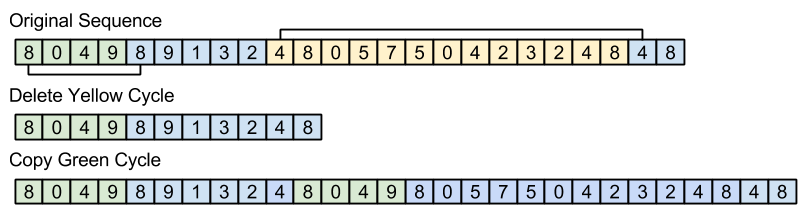}
\caption{Recombination mechanisms of copying or deleting a cycle.  Two possible cycles are selected from the original sequence.  The yellow cycle is deleted in the first recombination and the green sequence is copied into a location it fits in the second recombination.}
\label{fig:recomb}
\end{center}
\end{figure}

When recombination occurs all cycles in the genome are found (this can be done in a single pass over the genome).  A cycle occurs when the path returns to a site it has been to before.  Assuming at least one cycle exists (there are often many) we select one uniformly.  If a cycle deletion has been selected we delete the cycle.  If a cycle copy has been selected we find all suitable locations to insert the cycle and select one uniformly.  We then add the cycle to that location.  A cycle that starts and ends at site $i$ can replace any occurrence of site $i$ in the genome.  See Figure~\ref{fig:recomb} for examples of the recombination mechanisms.  It is important to note that there is no maximum cycle length enforced in our simulation at this time.  This means the cycle copy can result in a very large copy (in the most extreme case this doubles the size of the genome) or a very large deletion (in the most extreme case this deletes all but one gene).  When the whole genome is copied this is called genome duplication \citep{MDRC05}. 

During sexual reproduction we also have the likelihood of sexual recombination.  The recombination mechanism first computes the maximum number of crossover positions between the two paths by computing the longest common subsequence (similar to synapsis).  Every entry in the longest common subsequence is a site that the two sequences have in common.  We use these sites as crossover locations when creating an offspring genome.  You can imagine lining up the genomes so these sites are next to each other and copying the adjoining sequences from either the mother or father sequence.

\begin{figure}[!t]
\begin{center}
\includegraphics[width = 0.5\textwidth]{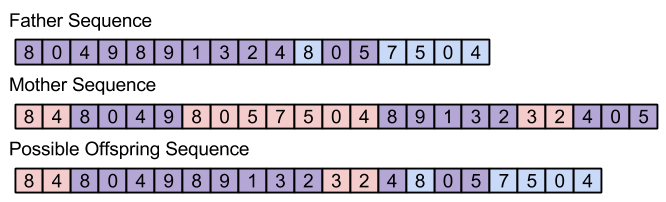}
\caption{Sexual recombination example.  The purple genes indicate the genes on the longest common subsequence.  The offspring sequence always includes the longest common subsequence as well as randomly selected adjoining sequences from its parents.}
\label{fig:sexrecomb}
\end{center}
\end{figure}

More formally the offspring genome is created first by selecting one of the two parents to begin copying from.  We copy genes from this agent to the offspring agent until we reach the first crossover location.  Whenever we reach a crossover location we flip a coin to determine which parent to read from next.  We then copy the section of genes from the newly selected agent until we again reach the next crossover location or the end of the genome.  We repeat until there are no more crossover locations.  Even when there is a lot of overlap between genomes (see Figure~\ref{fig:sexrecomb} as an example) the genes of each parent may have a different allele (i.e. a different strategy at that site).  Therefore, which parent we copy the crossover genes from also affects the creation of the offspring's genome.

After carrying out sexual recombination the new genome has probabilities for mutation or recombination with itself as well.  A new agent is born on the next day into the site where the parents conceived it.  There is the chance that a mutation or recombination has lead to a deletion that left the agent with no strategies given its birth location.  When this happens the agent is considered nonviable and dies upon birth.

\section{Observations}

Our agent-based model results in very rich emergent genetic dynamics and social structure.  This includes the evidence for genotype-phenotype divergence (e.g. phenotypic plasticity), non-coding regions of the genome \citep{H00}, pseudogenes \citep{MSRM00}, genetic drift \citep{HC97}, gene and genome duplication \citep{MDRC05}, and degrees of robustness and evolvability of the genome \citep{W12}.  There are also interesting population level dynamics emergent in our simulation.  These include the social organizations of herding \citep{R87}, natal philopatry and assortative mating \citep{JBK13} that emerge around sexual reproduction reported in \cite{MC15} as well as population migration, and sympatric/allopatric specialization \citep{V01}.  We do not have speciation explicitly built into our model, though genetic divergence of populations is observed even in overlapping territories.    

Analytically, we can argue how these features have emerged in our model.  For example, consider the divergence of the phenotype from the genotype.  One dimension of divergence is phenotypic plasticity, the ability to alter the phenotype in response to environmental features.  Two agents in our model that have an identical genotype may occupy different locations in the environment and thus select different phenotypes for the day. The other dimension of divergence occurs as phenotypically similar populations drift apart genotypically.  Two agents in our model with identical phenotypes must have some genetic similarity since the overlapping sequence must occur in both of their genomes.  However, this need be the only genetic similarities between agents.  When additions or deletions occur in regions of the agent's genome that are not used day-to-day these mutations will not affect the phenotype but will contribute to genetic differences.  Overtime these can accumulate into large genetic differences between agents that have the same phenotype (see below).

\begin{figure}[!t]
\begin{center}
\includegraphics[width = 0.5\textwidth]{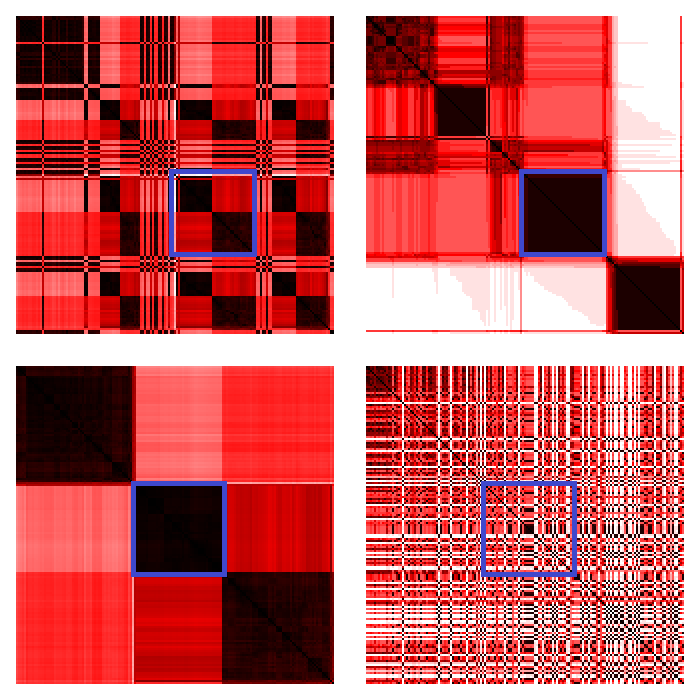}
\caption{Genotype-phenotype divergence.  Genotype (left column) and phenotype (right column) difference heat maps for an evolved population of 159 agents.  Black indicates similarity, white difference, and shades of red in between.  Agents are clustered by phenotype (top row) or genotype (bottom row).  The region highlighted in purple in the top row shows a phenotypically similar herd consisting of agents with different genotypes.  The region highlighted in purple in the bottom row shows a population of genetically similar agents with different phenotypes.}
\label{fig:genopheno}
\end{center}
\end{figure}

Evidence for these behaviors can be gathered based on a few advantages of our model.  First since genomes in our model are sequences we can use standard genome comparison methods like the Levenshtein edit-distance \citep{L66} to calculate mutation distance between agents.  A relevant phenotype of our agents is the sequence of sites they visit over the day.  Since the phenotypes of the agents are also sequences, they can be compared in the same way.  Figure~\ref{fig:genopheno} shows one population of 159 agents after evolving for 10000 days.  In the figure we use four heat maps to represent the variation in genotypes and phenotypes.  In each heat map the the rows and columns represent agents (in the same order) and the intersections have a color indicating how different the two agents are (genotypically or phenotypically).  We use single link hierarchical clustering to group our agents together based on their phenotypes and genotypes.  We see that when we cluster agents based on phenotype we get collections of agents we call herds in \cite{MC15}.  These can be identified as dark squares along the diagonal of the top right heat map in Figure~\ref{fig:genopheno}.  One herd has been indicated in purple.  When we look at the same group of agents' genotypes (top left heat map) we find that though there is some similarity, there are clearly at least two different genetic groups in this herd.  When we cluster the agents based on genotype instead (bottom left heat map) we see three clear genetic groups.  One is indicated in purple.  It is not hard to see that these groups consist of agents with widely different phenotypes (bottom right heat map).    

We know that the regions of DNA serve different roles roughly grouped as coding and non-coding regions.  Coding regions code for functional proteins whereas non-coding regions range from coding for other functional elements like RNA to a catch all ``junk DNA'' for regions that have no known function.  Some non-coding sequences called \emph{pseudogenes} are either genes that have lost their ability to be expressed or have otherwise ceased expression.  The identifiers of pseudogenes are their homology to other genes and their lack of expression.  Applying this definition to our model we can suggest that the regions of the genome that are not used by the agents are also homologous to regions that are used at least in so far as they potentially encode a phenotype.  So for this discussion we'll consider the region of the genome used by the agents day-to-day to be the \emph{expressed gene sequence} and the region that is not used to be the \emph{psuedogene sequence}.  

An important factor of the pseudogene sequences in our model is that they are neutral to selection but have potential functionality or even sub-functionality \citep{Q10, GO12}.  Mutations in the pseudogene sequence are commonly neutral because these regions are not used.  Assuming that mutations occur at a steady rate, then these neutral mutations can accumulate.  This process is called genetic drift.  Since the expressed gene sequence is rarely more than twenty genes in our model we can see genetic drift in the large pseudogene sequence.

\begin{figure}[!t]
\begin{center}
\includegraphics[width =0.5\textwidth]{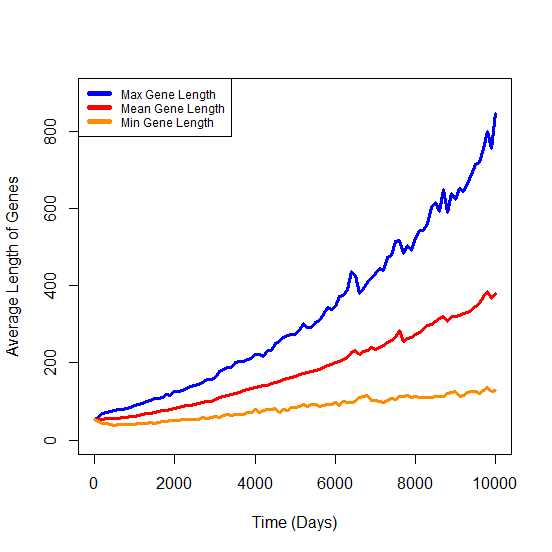}
\caption{Genome length over time.  Maximum, mean and minimum genome length averaged over hundreds of independent runs.}
\label{fig:length}
\end{center}
\end{figure}

This drift occurs in two ways.  First, the alleles in the pseudogene sequence will mutate (though these changes are not reflected in the phenotype and thus are not selected for or against).  As a result there is no trend towards optimization in the alleles of the pseudogene sequence, whereas we do see such optimization in the expressed gene sequence.  Second, additions and deletions in the psudogene sequence also have no effect on the phenotype.  As a result the length of the genome drifts due to these additions and deletions.  Recombinant additions (cycle copying) are unbounded in length so this can cause the length of the genome to become quite long over several generations by copying regions of the pseudogene sequence.  Figure~\ref{fig:length} shows the trends in genome length over time.  This data represents the average over hundreds of runs.  Both average length and variation in length increase over time.  This phenomena maybe caused by genetic drift or gene sequence duplication (see below).

The cycle copying recombination can play a different role when copying a region of the expressed gene sequence of the genome.  All cycle copying recombination can be seen as a type of gene duplication because the duplication of what we are calling genes (we are copying the specific alleles in these cases).  When this occurs in the expressed gene sequence we can also consider this operation a gene sequence duplication.  In particular, consider a special case in which the entire expressed gene sequence is duplicated.  Assuming this is the first time this recombination occurs in our agent, then it has two identical locally optimal gene sequences.  When daily activity selection occurs one of the two identical regions will be selected.  The other sequence will be part of the pseudogene sequence because it is not expressed.  This will not result in any phenotypic difference between the child agent and the parent agent that had only one such region.  Nonetheless, consider the difference allelic (or other) mutation will have on the parent and child.  The parent, with only one optimal region, has a high risk attached to mutating the region of use.  If the alleles in that region mutate they more often than not become less efficient.  Such mutations will likely mean the new agent is less fit.  This result will likely have a negative impact on the reproductive success of the agent.

Consider now the child agent with two such regions.  Mutation that occurs to one will not affect the other.  Thus if a negative allelic mutation occurs on one of the regions but not the other the agent is insulated from this mutation.  Its phenotype is unaltered as it can use the non-mutated  region.  Being insulated from negative mutations is a form of genetic robustness.  With this added robustness, the genome of the child agent is now able to explore the allelic space with less risk.  The chance of a positive mutation affecting the phenotype has doubled (it can occur in either region) but the chance of a negative mutation affecting the phenotype has halved (it has to occur in both regions).  This is a step toward what is called greater evolvability.  The gene sequence duplication we describe here results in greater robustness and evolvability.  It can also occur more than once (and in our model can potentially increase the number of regions exponentially per duplication).  Each duplication further impacts robustness and evolvability. 

\section{Extending our Model}

The genetic and population level dynamics we have encountered in our model, which may be among others we have not yet identified, are due to the commitment to emergent dynamics in as many aspects of our model as possible.  For instance the genetic dynamics discussed here are largely due to evolving a \emph{sequence of genes} with a realistic copying mechanism as is the case with DNA.  Even though this mechanism plays a major role, the structured genome could not exist without the structured environment.  Further the dynamics are greatly limited if we do not have genotype-phenotype divergence or metabolic systems \citep{MC14}.  We have tested a mechanism in which the whole genome is used by the agent (it travels back and forth along the genome) which causes convergence on short optimal genomes but is also less dynamic.

Too often steps are made to simplify the model computationally.  This typically means reducing one or many of these genetic sub-systems to linear dynamics.  This then removes possibilities for synergistic emergent dynamics among these systems.  Even so, there are many aspects of our model that still rely on simple rules and linear systems.  A short list of future improvements we would like to make are adding dynamism to the structure of the environment, increasing the genetic and environmental interaction, developing the energetic metabolism of our agents, and adding more genetic copying and deleting mechanics. 

The sequential properties of paths in a network allow for natural mechanisms of mutation and recombination to be employed.  These in turn lead to dynamics similar to those observed in nature and alow for the application of many biological concepts through analogy to our artificial agents.  Our agents make use of a foraging analogy \citep{M00} but evolving paths in a network can be used for other ends including optimization and solving other network-based problems in complex systems.  We encourage others to investigate the richness of this model.

In the short term, we plan to explore in more details the nature of the genetic dynamics displayed by our model.  The mechanisms of genetic drift and gene duplication deserve further study.  At present we are developing means of detecting and tracking events of recombinant duplication.  We are also currently developing metrics of robustness and evolvability for our agents that are independent of the gene sequence duplication.  We can use these to determine the impact of gene sequence duplication on the robustness and evolvability of the genomes of our agents.  Further we have observed allopatric and sympatric specialization of populations in different regions.  Simple examples of this (see \cite{MC15}) are the result of migration and separation in space (allopatric) or assortative mating (sympatric).

The goal of our long term research is to understand the origins of social behavior and social learning in artificial and biological populations.  We feel it is critical to these goals that the manner of social interaction in our models is not scripted as is common in multi-agent systems and agent based models in the social sciences.  Instead we desire a model in which social behaviors emerge under selection from underlying non-social forces.  Our current model demonstrates this first stage of development.  

The next major step is to allow those emergent social behaviors to create a new domain of interaction that can lead to higher level emergence.  We call this a cascade of emergence.  Complex life and especially the social behavior of humans is due to such a cascade.  To accomplish this we are developing coupled emergent systems for individual and social learning.  These are modeled off of our earlier work in \cite{MC14}.

\footnotesize
\bibliographystyle{apalike}
\bibliography{example}

\begin{thebibliography}{}

\bibitem[Antonioni et~al., 2014]{ABT14}
Antonioni, A., Bullock, S., and Tomassini, M. (2014).
\newblock Reds: an energy-constrained spatial social network model.
\newblock In {\em The Fourteenth Conference on the Synthesis and Simulation of
  Living Systems}, volume~14. MIT Press.

\bibitem[Bonabeau, 2002]{B02}
Bonabeau, E. (2002).
\newblock Agent-based modeling: Methods and techniques for simulating human
  systems.
\newblock {\em Proceedings of the National Academy of Sciences}, 99(suppl
  3):7280--7287.

\bibitem[Chalmers, 2006]{C06}
Chalmers, D.~J. (2006).
\newblock Strong and weak emergence.
\newblock {\em The reemergence of emergence}, pages 244--256.

\bibitem[Conrad and Rizki, 1989]{CR89}
Conrad, M. and Rizki, M.~M. (1989).
\newblock The artificial worlds approach to emergent evolution.
\newblock {\em BioSystems}, 23(2):247--258.

\bibitem[Dennett, 1995]{D95}
Dennett, D. (1995).
\newblock {\em Darwin's dangerous idea}.
\newblock New York: Simon and Schuster.

\bibitem[Ellis, 2006]{E06}
Ellis, G.~F. (2006).
\newblock On the nature of emergent reality.
\newblock {\em The re-emergence of emergence}, pages 79--107.

\bibitem[Epstein, 1999]{E99}
Epstein, J.~M. (1999).
\newblock Agent-based computational models and generative social science.
\newblock {\em Generative Social Science: Studies in Agent-Based Computational
  Modeling}, 4(5):4--46.

\bibitem[Fodor, 1974]{F74}
Fodor, J.~A. (1974).
\newblock Special sciences (or: the disunity of science as a working
  hypothesis).
\newblock {\em Synthese}, 28(2):97--115.

\bibitem[Gleick, 1987]{G87}
Gleick, J. (1987).
\newblock {\em Chaos: Making of a new science}.
\newblock New York: Penguin.

\bibitem[Guillaume and Otto, 2012]{GO12}
Guillaume, F. and Otto, S.~P. (2012).
\newblock Gene functional trade-offs and the evolution of pleiotropy.
\newblock {\em Genetics}, 192(4):1389--1409.

\bibitem[Hansen and Wagner, 2001]{HW01}
Hansen, T.~F. and Wagner, G.~P. (2001).
\newblock Modeling genetic architecture: a multilinear theory of gene
  interaction.
\newblock {\em Theoretical population biology}, 59(1):61--86.

\bibitem[Hardison, 2000]{H00}
Hardison, R.~C. (2000).
\newblock Conserved noncoding sequences are reliable guides to regulatory
  elements.
\newblock {\em Trends in Genetics}, 16(9):369--372.

\bibitem[Hartl and Clark, 1997]{HC97}
Hartl, D.~L. and Clark, A.~G. (1997).
\newblock {\em Principles of population genetics}.
\newblock Sinauer associates Sunderland.

\bibitem[Heath et~al., 2009]{HHC09}
Heath, B., Hill, R., and Ciarallo, F. (2009).
\newblock A survey of agent-based modeling practices (january 1998 to july
  2008).
\newblock {\em Journal of Artificial Societies and Social Simulation}, 12(4):9.

\bibitem[Inada, 2001]{I01}
Inada, Y. (2001).
\newblock Steering mechanism of fish schools.
\newblock {\em Complexity international}, 8:1--9.

\bibitem[Jiang et~al., 2013]{JBK13}
Jiang, Y., Bolnick, D.~I., and Kirkpatrick, M. (2013).
\newblock Assortative mating in animals.
\newblock {\em The American Naturalist}, 181(6):E125--E138.

\bibitem[Johnson, 2002]{J02}
Johnson, S. (2002).
\newblock {\em Emergence: The connected lives of ants, brains, cities, and
  software}.
\newblock Simon and Schuster.

\bibitem[Kim, 1992]{K92}
Kim, J. (1992).
\newblock Multiple realization and the metaphysics of reduction.
\newblock {\em Philosophy and Phenomenological Research}, pages 1--26.

\bibitem[Levenshtein, 1966]{L66}
Levenshtein, V.~I. (1966).
\newblock Binary codes capable of correcting deletions, insertions, and
  reversals.
\newblock In {\em Soviet physics doklady}, volume~10, pages 707--710.

\bibitem[Lov{\'a}sz, 1993]{l93}
Lov{\'a}sz, L. (1993).
\newblock Random walks on graphs: A survey.
\newblock {\em Combinatorics, Paul erdos is eighty}, 2(1):1--46.

\bibitem[Macy and Willer, 2002]{MW02}
Macy, M.~W. and Willer, R. (2002).
\newblock From factors to actors: Computational sociology and agent-based
  modeling.
\newblock {\em Annual review of sociology}, pages 143--166.

\bibitem[Maere et~al., 2005]{MDRC05}
Maere, S., De~Bodt, S., Raes, J., Casneuf, T., Van~Montagu, M., Kuiper, M., and
  Van~de Peer, Y. (2005).
\newblock Modeling gene and genome duplications in eukaryotes.
\newblock {\em Proceedings of the National Academy of Sciences of the United
  States of America}, 102(15):5454--5459.

\bibitem[Marriott and Chebib, 2014]{MC14}
Marriott, C. and Chebib, J. (2014).
\newblock The effect of social learning on individual learning and evolution.
\newblock In {\em The Fourteenth Conference on the Synthesis and Simulation of
  Living Systems}, volume~14, pages 736--743. MIT Press.

\bibitem[Marriott and Chebib, 2015]{MC15}
Marriott, C. and Chebib, J. (2015).
\newblock Finding a mate with no social skills.
\newblock In {\em Proceedings of the 2015 conference on Genetic and
  Evolutionary Computation}. ACM.

\bibitem[Melo and Marroig, 2015]{MM15}
Melo, D. and Marroig, G. (2015).
\newblock Directional selection can drive the evolution of modularity in
  complex traits.
\newblock {\em Proceedings of the National Academy of Sciences},
  112(2):470--475.

\bibitem[Mighell et~al., 2000]{MSRM00}
Mighell, A., Smith, N., Robinson, P., and Markham, A. (2000).
\newblock Vertebrate pseudogenes.
\newblock {\em Febs Letters}, 468(2):109--114.

\bibitem[Miller et~al., 2008]{MP08}
Miller, J.~H., Page, S.~E., and LeBaron, B. (2008).
\newblock {\em Complex adaptive systems: an introduction to computational
  models of social life}.
\newblock Princeton: Princeton University Press.

\bibitem[Mobus, 2000]{M00}
Mobus, G. (2000).
\newblock Foraging search: Prototypical intelligence.
\newblock In {\em Proceedings of the Conference of the American Institute of
  Physics}, pages 592--605. Institute of Physics Publishing.

\bibitem[Morowitz, 2002]{M02}
Morowitz, H.~J. (2002).
\newblock {\em The emergence of everything: How the world became complex}.
\newblock Oxford University Press.

\bibitem[Nowak and Sigmund, 2004]{NS04}
Nowak, M.~A. and Sigmund, K. (2004).
\newblock Evolutionary dynamics of biological games.
\newblock {\em science}, 303(5659):793--799.

\bibitem[Olfati-Saber, 2006]{OS06}
Olfati-Saber, R. (2006).
\newblock Flocking for multi-agent dynamic systems: Algorithms and theory.
\newblock {\em Automatic Control, IEEE Transactions on}, 51(3):401--420.

\bibitem[Penrose, 2003]{P03}
Penrose, M. (2003).
\newblock {\em Random geometric graphs}.
\newblock Oxford University Press Oxford.

\bibitem[Polly, 2008]{P08}
Polly, P.~D. (2008).
\newblock Developmental dynamics and g-matrices: Can morphometric spaces be
  used to model phenotypic evolution?
\newblock {\em Evolutionary biology}, 35(2):83--96.

\bibitem[Qian et~al., 2010]{Q10}
Qian, W.~F., Liao, B.~Y., Chang, A. Y.~F., and Zhang, J.~Z. (2010).
\newblock Maintenance of duplicate genes and their functional redundancy by
  reduced expression.
\newblock {\em Trends Genet.}, 26(10):425--430.

\bibitem[Reynolds, 1987]{R87}
Reynolds, C.~W. (1987).
\newblock Flocks, herds and schools: A distributed behavioral model.
\newblock {\em ACM Siggraph Computer Graphics}, 21(4):25--34.

\bibitem[Szathmary and Maynard~Smith, 2004]{SM04}
Szathmary, E. and Maynard~Smith, J. (2004).
\newblock {\em The major transitions in evolution}.
\newblock Oxford University Press.

\bibitem[Via, 2001]{V01}
Via, S. (2001).
\newblock Sympatric speciation in animals: the ugly duckling grows up.
\newblock {\em Trends in Ecology \& Evolution}, 16(7):381--390.

\bibitem[Wagner, 2012]{W12}
Wagner, A. (2012).
\newblock The role of robustness in phenotypic adaptation and innovation.
\newblock {\em Proceedings of the Royal Society B: Biological Sciences},
  279(1732):1249--1258.

\bibitem[Wolfe and Goldberg, 2000]{WG00}
Wolfe, M.~F. and Goldberg, R. (2000).
\newblock {\em Rube Goldberg: Inventions!}
\newblock Simon and Schuster.

\end{thebibliography}

\end{document}